\documentstyle[aps,epsfig]{revtex}
\begin{document}
\title{Semiclassical description of multiple giant dipole resonance excitation and
decay$^{*}$}
\author{B.V.~Carlson$^{1,**}$, M.S.~Hussein$^{2,**}$ and A.F.R.~de Toledo Piza$^2$}
\address{$^1$Departamento de F\'{\i}sica do Instituto Tecnol\'ogico de\\
Aeron\'autica - CTA\\
12228-900 S\~ao Jos\'e dos Campos, SP, Brazil}
\address{$^2$Instituto de F\'{\i}sica, Universidade de S\~ao Paulo,\\
C.P. 66318, S\~ao Paulo, 05315-970, Brazil}
\maketitle

\begin{abstract}
A semiclassical description of multiple giant resonance excitation and decay
that incorporates incoherent fluctuation contributions of the Brink-Axel type
is developed. Numerical calculations show that the incoherent contributions
are important at low to intermediate bombarding energies.
\end{abstract}

%\vspace{8cm}

\begin{quote}
$*$~Supported in part by FAPESP.\newline
$**$Supported in part by CNPq.
\end{quote}

%\newpage

We wish to develop a schematic semiclassical description of the
collective-statistical theory of multiple giant dipole resonance excitation
and decay given in Refs.~\onlinecite{ccchp} and \onlinecite{ccchp2}.
To do this, we use the formalism developed in Ref.~\onlinecite{ko} to
extend the semiclassical description of collective multiple giant resonance
excitation given in Ref.~\onlinecite{canto}.
This development is important for the understanding of the recent observation
and analysis of the heavy-ion Coulomb excitation of the Double Giant Dipole
Resonance reported in Refs.~\onlinecite{emling1} and \onlinecite{emling2}.

As in Ref.~\onlinecite{ccchp}, we wish to take into account the contribution
to the cross section of collective nuclear excitations that occur both before
and after the statistical decay of earlier collective excitations. We
thus label the states of the nucleus with both a
collective index $n$, denoting the number of collective dipole phonons, and
a statistical one $s$, denoting the number of collective phonons that have
decayed into the incoherent background. The class of states denoted by the
pair of indices $n$ and $s$ thus possesses $n$ phonons of collective
excitation and an incoherent background excitation obtained through the
decay of another $s$ phonons. We will represent this class of states by a
single state. In the limit of harmonic phonons, this state would have an
excitation energy of $(n+s)\varepsilon _{d}$ and a width of $n\Gamma _{d}$,
where $\varepsilon _{d}$ is the energy of the giant dipole resonance and $%
\Gamma _{d}$ is its spreading width. We will neglect contributions of the
escape widths, as these are extremely small when compared to those of the
spreading widths of the systems of interest here.\cite{chp}

Although the collective excitation of the nucleus is a coherent
process, its statistical decay is an incoherent one. The time evolution of
the system thus possesses both coherent and incoherent aspects,
making a density matrix formulation necessary. We consider the
time evolution of the density matrix elements $\rho _{ns,n^{\prime
}s^{\prime }}(t)$. The diagonal element $\rho _{ns,ns}(t)$ represents the
instantaneous occupation probability of the state with $n$ collective
phonons and a statistical background equivalent to $s$ phonons.

Following Ref. \cite{ko}, we can put the time-evolution equation of the
semiclassical density matrix into the form 
\begin{eqnarray}
\hbar \frac{\partial \rho _{ns,n^{\prime }s^{\prime }}}{\partial t}
&=&-i\sum_{m}\left\{ \left( \varepsilon _{n}+\varepsilon _{s}\right) \delta
_{nm}+V_{nm}(t)\right\} \rho _{ms,n^{\prime }s^{\prime }}  \nonumber \\
&&+i\sum_{m}\rho _{ns,ms^{\prime }}\left\{ \left( \varepsilon _{n^{\prime
}}+\varepsilon _{s^{\prime }}\right) \delta _{mn^{\prime }}+V_{mn^{\prime
}}(t)\right\} \\
&&-\frac{(\Gamma _{ns}+\Gamma _{n^{\prime }s^{\prime }})}{2}\rho
_{nsn^{\prime }s^{\prime }}+\delta _{nn^{\prime }}\delta _{ss^{\prime
}}\sum_{m,r}\Gamma _{ns\leftarrow mr}\rho _{mr,mr},  \nonumber
\end{eqnarray}
The terms in the first two lines on the right-hand side induce the coherent
contribution to the evolution. This is given in terms of the
(diagonal) collective and
statistical contributions to the excitation energy, $\varepsilon _{n}$ and $%
\varepsilon _{s}$, and of the interaction $V$, which couples the states
through collective excitation alone. The two terms on the third line describe,
respectively, the loss of probability due to incoherent, statistical
transitions out of the state and the gain of probability due to statistical
transitions from the other states. The partial gain widths 
$\Gamma _{ns\leftarrow mr}$ are such that 
\begin{equation}
\sum_{n,s}\Gamma _{ns\leftarrow mr}=\Gamma _{mr}.
\end{equation}
This condition simply states that the sum of the partial widths for
probability transfer from any one state to all others must be equal to the
total width for probability loss from the initial state. This guarantees
probability conservation during the evolution of the system (assuming, of
course, that $V$ is Hermitian).

We will assume that the initial population is in the ground state. The
initial conditions for which the equation will be solved are then
\begin{equation}
\rho _{ns,n^{\prime}s^{\prime}}(t\rightarrow -\infty )\rightarrow 
\delta _{nn^{\prime}}\delta _{n0}\delta _{ss^{\prime}}\delta _{s0} (1-T(b)),
\end{equation}
where $T(b)$ is an impact-parameter dependent transmission coefficient
that takes into account the probability of projectile-target interactions
more complex than those being discussed here. We approximate the
transmission coefficient as 
\begin{equation}
T(b)=\frac{1}{1+\exp((R-b)/a))},\label{transcof}
\end{equation}
where we take the strong-interaction radius to be
$R=1.23(A_P^{1/3}+A_T^{1/3})$ fm and the diffusivity to be $a= 0.75$ fm,
with $A_P$ and $A_T$ the projectile and target mass numbers,
respectively.

As the only coherent coupling in the time-evolution equation is
through the collective interaction $V$, which couples only
collective states having the same statistical index $s$, we conclude that
the density matrix will remain diagonal in the statistical index $s$ at all
times,
\begin{equation}
\rho _{ns,n^{\prime}s^{\prime}}(t)=\delta _{ss^{\prime}}\rho
_{ns,n^{\prime}s}(t).
\end{equation}
The density matrix thus reduces to a separate density submatrix for each
value of the statistical index, with the coupling between these submatrices,
through the gain and loss terms, being completely incoherent.

It is convenient to explicitly take into account the time dependence due to
the collective excitation energy. To do this, we define a modified density
matrix, which will have the same diagonal matrix elements as the original
one, as
\begin{equation}
\rho _{nn^{\prime}}^{s}(t)=\exp \left[ -i(\varepsilon _{n}-\varepsilon
_{n^{\prime}})t/\hbar\right] \rho _{ns,n^{\prime}s}(t).
\end{equation}
The time evolution equation then reduces to the form
\begin{eqnarray}
\hbar\frac{\partial \rho _{nn^{\prime}}^{s}}{\partial t} &=&-i\sum_{m}\left( 
\tilde{V}_{nm}(t)\rho _{mn^{\prime}}^{s}-\rho _{nm}^{s}\tilde{V}%
_{mn^{\prime}}(t)\right)  \label{evoleq} \\
&&-\frac{(\Gamma _{ns}+\Gamma _{n^{\prime}s})}{2}\rho
_{nn^{\prime}}^{s}+\delta _{nn^{\prime}}\sum_{r,m}\Gamma _{ns\leftarrow
mr}\rho _{mm}^{r},  \nonumber
\end{eqnarray}
in which the remaining contribution to the coherent evolution is due to 
$\tilde{V}$, where
\begin{equation}
\tilde{V}_{nn^{\prime }}(t)=\exp \left[ i(\varepsilon _{n}-\varepsilon
_{n^{\prime }})t/\hbar \right] V_{nn^{\prime }}(t).
\end{equation}
The second line of Eq.(\ref{evoleq}) contains the incoherent contributions of
the statistical loss and gain terms, respectively.

Assuming that the collective excited states are harmonic $n$-phonon giant
dipole states, the interaction matrix elements can be written as
\begin{equation}
\tilde{V}_{nn^{\prime}}(t)=\left( \exp \left[ i\varepsilon _{d}t/\hbar
\right] \sqrt{n}\delta _{n^{\prime},n-1}+\exp \left[ -i\varepsilon _{d}t/
\hbar \right] \sqrt{n+1}\delta _{n^{\prime},n+1}\right) V_{01}(t)
\end{equation}
where $\varepsilon _{d}$ is the excitation energy of the giant dipole
resonance and $V_{01}(t)$ is the semiclassical matrix element coupling the
ground state to the giant resonance, which we take to have the simple form 
\begin{equation}
V_{01}(t)=V_{0}\frac{\left( b_{\min }/b \right) ^{2}}{1+(\gamma vt/b)^{2}},
\end{equation}
as given in Ref. \cite{canto}. As is done there, we neglect the spin
degeneracies and magnetic multiplicities of the giant resonance
states and approximate the projectile-target relative motion as a
straight line.

The decay widths in the case of harmonic phonons can be approximated as 
\begin{equation}
\Gamma _{ns}=n\Gamma _{d},
\end{equation}
where $\Gamma _{d}$ is the spreading width of the giant dipole resonance. We
have neglected the contribution to the width of the hot statistical
background of states since, at the low temperatures involved here, the decay
widths of the hot Brink-Axel resonances are very similar to those of the
cold ones.

According to the convention we have adopted for labeling states, the
statistical index denotes the number of collective phonons that have decayed
to the incoherent statistical background. The decay of the $n$-phonon
$s$-background state thus transfers its occupation probability to the
$(n-1)$-phonon $(s+1)$-background state. The form of the gain terms reflects
this fact, 
\begin{equation}
\Gamma _{ns\leftarrow mr}=\delta _{s,r+1}\delta _{n,m-1}\Gamma _{mr}=\delta
_{s,r+1}\delta _{n,m-1}m\Gamma _{d}.
\end{equation}

We observe, from the form of the time-evolution equation, that all states
will eventually decay to the states containing no collective excitations.
We thus have for the asymptotic occupation probabilities,
\begin{equation}
\rho _{00}^{s}(t\rightarrow \infty )\rightarrow P(n=0,s),
\end{equation}
with the $P(n=0,s)$ being defined by this limit. All other matrix elements
tend to zero,
\begin{equation}
\rho _{nm}^{s}(t\rightarrow \infty )\rightarrow 0\hspace{1cm}n,m\neq 0.
\end{equation}
Conservation of probability requires that
\begin{equation}
\sum_{s}P(n=0,s)=1-T(b),
\end{equation}
where $T(b)$ is the transmission coefficient of Eq.(\ref{transcof}).

Although the states containing collective phonons are asymptotically
depopulated, we can still obtain an estimate of the probability that passes
through them by calculating the probability that decays out of them. We thus
define for these states
\begin{equation}
P(n\neq 0,s)\equiv \Gamma _{ns}\int_{-\infty }^{\infty }dt\,\rho
_{nn}^{s}(t).
\end{equation}
We note that this is only an estimate of the probability that has passed
through each state, as it takes into account only that part of the
probability that decays incoherently. It does not include the fraction of
the probability that was transferred coherently (through the action of $V$)
to other states.

Finally, we define a cross section $\sigma _{ns}$ for each state by
integrating its probability $P(n,s)$ over the implicit dependence on the
impact parameter,
\begin{equation}
\sigma _{ns}\equiv 2\pi \int_{b_{min}}^{\infty }b\,db\,P(n,s).
\end{equation}
At extremely low energies, the lower limit of the integral over impact
parameter,
$b_{min}$, is determined by the classical point of closest approach of the
Coulomb interaction. When the energy is sufficiently high to surpass the
Coulomb barrier, the transmission coefficient $T(b)$ cuts the integral
off at low values of the impact parameter.

We have performed calculations of multiple giant dipole resonance excitation
within the model for the system $^{208}$Pb + $^{208}$Pb in the projectile
energy range from 100 to 1000 Mev/nucleon. For the centroid and width
of the giant dipole resonance, we use values taken from a global systematic,
$\varepsilon_{d}=43.4\, A^{-0.215}$ MeV and
$\Gamma_d=0.3\, \varepsilon_{d}$,\cite{reffo} giving
$\varepsilon_{d}=13.8$ MeV and $\Gamma_d=4.1$ MeV,
slightly above the experimental values.

 We display in Fig.~1, as a function of the projectile energy, the coherent
$n$-phonon cross sections $\sigma_{n0}$ (dashed lines) and total $n$-phonon
cross sections $\sigma_{0n}$ (solid lines) obtained from the calculation.
The coherent cross sections $\sigma_{n0}$ describe the direct excitation
of the $n$-phonon states. These are the cross sections that result
from a typical calculation of multiple giant resonance excitation amplitudes. 
The total $n$-phonon cross sections $\sigma_{0n}$ account for all possible
$n$-phonon excitations, including those in which one or more of the
phonons decays incoherently before others are excited. The respective
coherent or total cross sections decrease by about an order of magnitude
for each additional phonon of excitation. The coherent
$n$-phonon cross sections increase monotonically with energy, as do
the total excitation cross sections for low phonon number. For the
cases of three or more phonons, the total $n$-phonon cross section
first decreases with the incident energy, but then turns and
increases like the other cross sections. 

Except for the 1-phonon case, the total $n$-phonon cross sections 
$\sigma_{0n}$ in Fig.~1 are clearly larger than the coherent cross
sections $\sigma_{n0}$. This can be readily understood by noting
that, although there is only one way a single phonon can be excited
and decay, alternative sequences of excitation and decay are
available when more than one phonon is involved.  As we have 
emphasized previously in the case of 2-phonons\cite{ccchp,ccchp2,chp},
the total cross section $\sigma_{02}$ contains both the coherent
2-phonon excitation, 2-phonon decay contribution $\sigma_{20}$ and an
incoherent contribution due to the excitation (and decay) of a second
phonon after the first phonon has decayed into the statistical background.
The apparent discrepancy in experimental double giant dipole
resonance cross sections\cite{emling1,emling2} can thus be explained by
arguing that what is observed is the total 2-phonon cross section
$\sigma_{02}$ and not just the coherent cross section $\sigma_{20}$.

The relative importance of the coherent excitation cross sections, 
$\sigma_{n0}$, compared to the total ones, $\sigma_{0n}$, can best
be seen by looking at their ratio, $\sigma_{0n}/\sigma_{n0}$, as shown
in Fig.~2 as a function of the projectile energy. We observe that the
total n-phonon cross sections $\sigma_{0n}$ are greatly enhanced relative to
the coherent cross sections $\sigma_{n0}$ at low energies. As the
energy increases, the relative enhancement decreases and
tends toward one. This trend can be explained by
comparing the time scale of the collision process to that of the
decay of a giant resonance into the statistical background. At low
bombarding energy, the collision occurs slowly relative to the decay
time of the resonance. Subsequent excitations then usually occur after
the previous ones have decayed and the cross sections for coherent
multiple excitation are small compared to the total ones. As the
energy increases, the collision time decreases and the time 
available for decay of a phonon before the excitation of another also
decreases. The relative importance of the incoherent
contributions to the n-phonon excitation cross section thus decreases
as does the relative enhancement of the total cross section over the
coherent one.

There has been a good deal of discussion in recent years related to
the fact that the observed width of the double giant dipole resonance
deviates from the harmonic value of twice the single giant resonance width.
As we have commented previously,\cite{ccchp} this is a natural result
of the incoherent contributions to the cross section, even when the
resonances themselves are purely harmonic. In fact, due to the energy
dependence of the incoherent contributions to the excitation cross
sections, we expect the effective $n$-phonon width to be energy dependent.
We can estimate the effective widths here by averaging the width over
the $n$-phonon cross sections, taking care to discount that part of
each cross section that results from the decay of the previous state in
the chain. That is, we take
\begin{eqnarray}
\Gamma_{eff,n}&=&\frac{\Gamma_{n0}\sigma_{n0}
              +\Gamma_{n-1,1}(\sigma_{n-1,1}-\sigma_{n0})+...
              +\Gamma_{1,n-1}(\sigma_{1,n-1}-\sigma_{2,n-2})}
{\sigma_{n0}+(\sigma_{n-1,1}-\sigma_{n0})+...+
(\sigma_{1,n-1}-\sigma_{2,n-2})} \nonumber \\
             &=&\Gamma_d\frac{\sigma_{n0}+\sigma_{n-1,1}+...+\sigma_{1,n-1}}
                             {\sigma_{1,n-1}}, \label{gambar}
\end{eqnarray}
where we have used the harmonic limit assumed in the calculations to obtain the
second expression. The results of this calculation for the system 
$^{208}$Pb + $^{208}$Pb, as a function of the projectile energy, are
shown in Fig.~3. We observe that the effective width of the
$n$-phonon excitation cross section is much smaller than the harmonic
value of $n\Gamma_d$ at low energy but approaches the harmonic value at
high energy. Such an increase in the effective width has been observed
experimentally in the case of the double giant dipole resonance.\cite{emling2}

Finally, in Fig.~4, we show the differential excitation cross
section that we obtain  for the system $^{208}$Pb + $^{208}$Pb at 640
MeV/nucleon as a function of the excitation energy. This was obtained
by summing Breit-Wigner expressions with the appropriate excitation
energy and width for each of the $n$-phonon cross sections, again taking
care to discount that part of each cross section that results from the
decay of the previous state in the chain. We show only the
contributions of the first three giant dipole resonances, as the
higher order ones are almost invisible even on our theoretical curve.
Only the first and second giant dipole resonances have been observed
experimentally.

In summary, the semiclassical calculations presented here allow
us to conclude that the collective-statistical theory of multiple giant
resonance excitation and decay provides
a theoretical basis for the energy-dependent enhancement of multiple
excitation cross sections and energy-dependent effective widths observed
experimentally.

\begin{figure}
\begin{center}
\epsfig{file=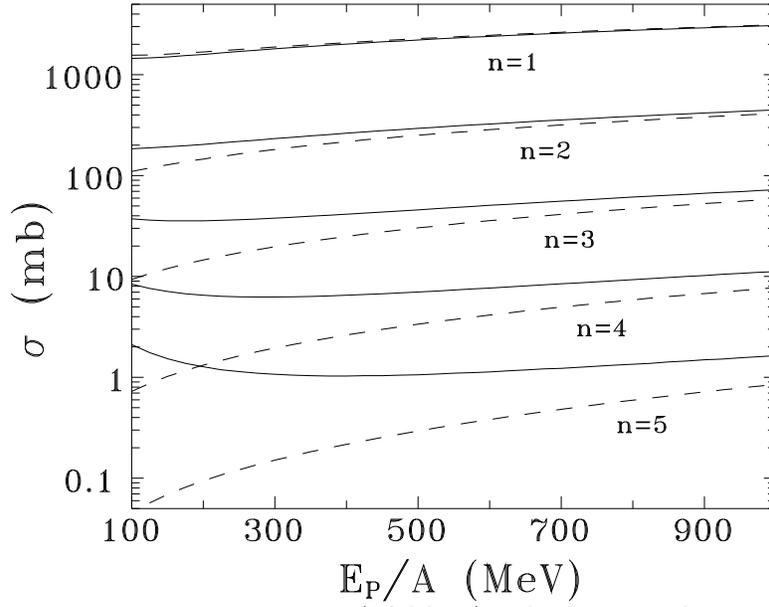,height=8cm}
\caption{ Total $n$-phonon excitation cross sections $\sigma_{0n}$
(solid lines) and coherent $n$-phonon excitation cross sections
$\sigma_{n0}$ (dashed lines) for the system $^{208}$Pb + $^{208}$Pb
as a function of the projectile energy.}
\end{center}
\label{fig1}
\end{figure}

\begin{figure}
\begin{center}
\epsfig{file=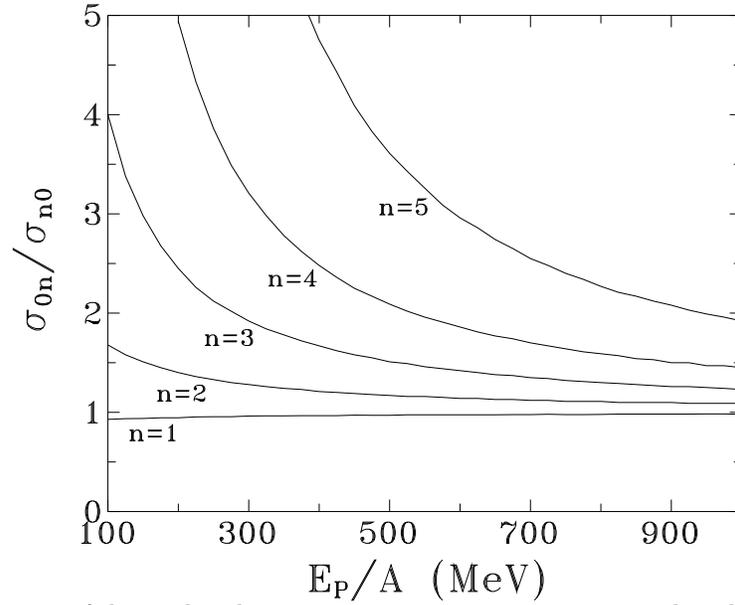,height=8cm}
\caption{ Relative enhancement of the total $n$-phonon excitation cross section
$\sigma_{0n}$ over the coherent excitation cross section $\sigma_{n0}$ for
the system $^{208}$Pb + $^{208}$Pb as a function of the projectile
energy.}
\end{center}
\label{fig2}
\end{figure}

\begin{figure}
\begin{center}
\epsfig{file=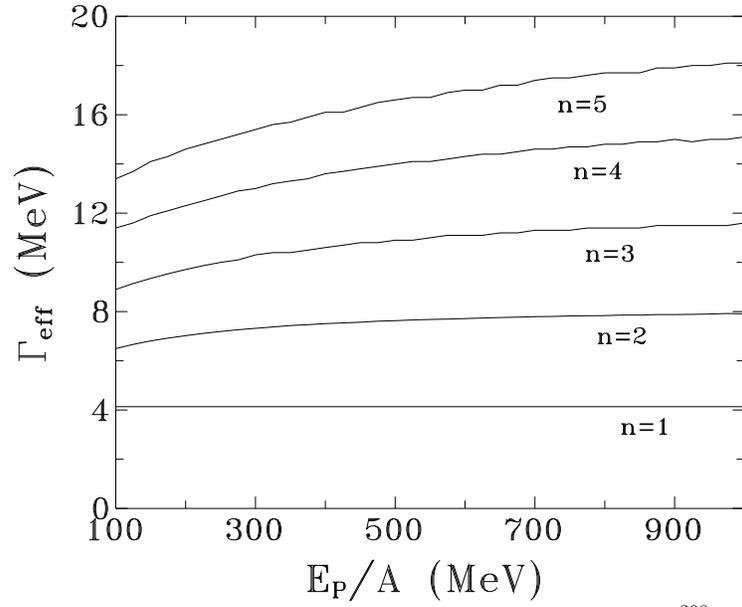,height=8cm}
\caption{ Effective widths of the first five multiple giant dipole resonances
of the system $^{208}$Pb + $^{208}$Pb as a function of the projectile
energy.}
\end{center}
\label{fig3}
\end{figure}

\begin{figure}
\begin{center}
\epsfig{file=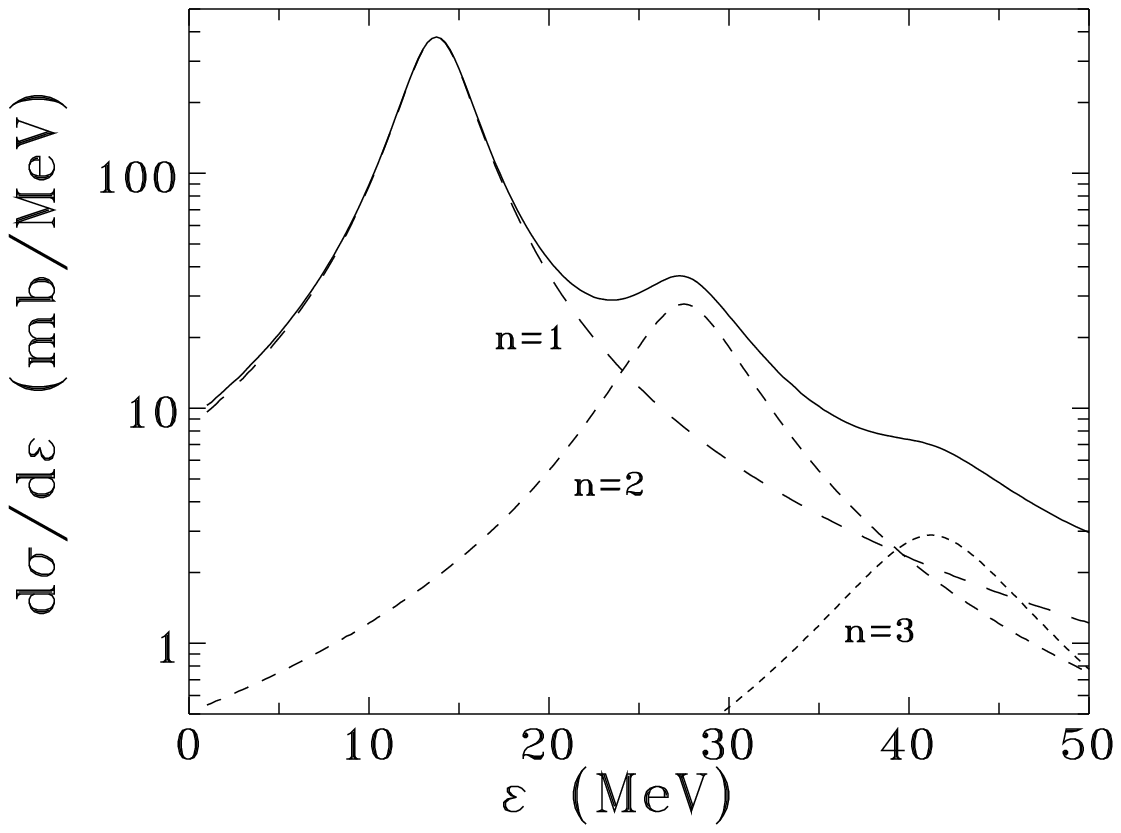,height=8cm}
\caption{ Theoretical multiple giant resonance differential excitation
cross section of $^{208}$Pb at a projectile energy of 640 MeV/nucleon.}
\end{center}
\label{fig4}
\end{figure}


\begin{references}
\bibitem{ccchp}   B.V. Carlson, L.F. Canto, S. Cruz-Barrios, M.S. Hussein,
and A.F.R. de Toledo Piza,``Multiphonon and ``hot''-phonon isovector
electric-dipole excitations'', submitted for publication.

\bibitem{ccchp2}  L.F. Canto, B.V. Carlson, S. Cruz-Barrios, M.S. Hussein,
and A.F.R. de Toledo Piza,``Fluctuation contributions to double giant dipole
resonance excitation cross sections'', to appear in Proceedings of the XX
Brazilian Nuclear Physics Workshop, World Scientific, Sept. 1997.

\bibitem{ko}  C.M. Ko, Z. Phys. A {\bf 286} (1978) 405.

\bibitem{canto}  L.F. Canto, A. Romanelli, M.S. Hussein and A.F.R. de Toledo
Piza, Phys. Rev. Lett. {\bf 72} (1994) 2147.

\bibitem{emling1} See, e.g., H. Emling, Prog. Part. Nucl. Phys. {\bf 33} (1994)
729.

\bibitem{emling2} K. Boretsky et al., Phys. Lett. {\bf B384} (1996) 30.

\bibitem{chp}  B.V. Carlson, M.S. Hussein, and A.F.R. de Toledo Piza, Phys.
Lett. {\bf B431} (1998) 249.

\bibitem{reffo} G.Reffo, private communication.

\end{references}
\end{document}